\documentclass[preprint,12pt]{aastex}

\newcommand{\degree}{\mbox{$^{\circ}$}}
\newcommand{\am}{\mbox{\arcmin}}
\newcommand{\as}{\mbox{\arcsec}}

\newcommand{\kms}{\mbox{km s$^{-1}$}}

\newcommand{\um}{$\mu$m}

\newcommand{\lsun}{\mbox{L$_\odot$}}
\newcommand{\lacc}{\mbox{$L_{acc}$}} 
\newcommand{\msun}{\mbox{M$_\odot$}}
\newcommand{\rsun}{\mbox{R$_\odot$}}

\newcommand{\lbol}{\mbox{$L_{bol}$}} 

\newcommand{\fsmm}{\mbox{$L_{smm}/L_{bol}$}} 
 
\def\lsun{$L_{\odot}$}
\def\um {$\mu$m}

\begin{document}


\title{\bf The $Spitzer$ c2d Survey of Nearby Dense Cores: 
Jet and Molecular Outflow Associated with a YSO in core A of L1251}
\author {Jeong-Eun Lee\altaffilmark{1},
Ho-Gyu Lee\altaffilmark{2},
Jong-Ho Shinn\altaffilmark{3},
Michael M. Dunham\altaffilmark{4},
Il-Suk Kim\altaffilmark{1},
Chang Hee Kim\altaffilmark{5},
Tyler L. Bourke\altaffilmark{6},
Neal J. Evans II\altaffilmark{4},
Yunhee Choi\altaffilmark{1}
}
\altaffiltext{1}{Department of Astronomy and Space Science, Astrophysical
Research Center for the Structure and Evolution of the Cosmos,
Sejong University, Seoul 143-747, Korea; jelee@sejong.ac.kr}
\altaffiltext{2}{Department of Astronomy, Graduate School of Science,
The University of Tokyo, Bunkyo-ku, Tokyo 113-0033, Japan;
hglee@astron.s.u-tokyo.ac.jp}
\altaffiltext{3}{Korea Astronomy and Space Science Institute, Daejeon, Korea, 305-348}
\altaffiltext{4}{Astronomy Department, The University of Texas at Austin,
1 University Station C1400, Austin, TX 78712-0259; mdunham@astro.as.utexas.edu,
nje@astro.as.utexas.edu}
\altaffiltext{5}{Department of Physics and Astronomy, Seoul National University, 
599 Gwanangno, Gwanak-gu, Seoul, 151-747, Korea}
\altaffiltext{6}{Smithsonian Astrophysical Observatory, 60 Garden Street, 
Cambridge, MA 02138; tbourke@cfa.harvard.edu}


\begin{abstract}
A long infrared jet has been discovered by the $Spitzer$ c2d legacy program in 
core A of L1251. It is associated with a very embedded Class 0 
object with an accretion luminosity of about 0.9 \lsun\ derived by 
radiative transfer 
model fitting to the observed SED.  Comparing the observed IRAC colors along 
the infrared jet with those calculated from a model of an admixture of gas 
with a power-law temperature distribution indicates that the jet is possibly 
created by a paraboloidal bow shock propagating into the 
ambient medium of $n(\rm H_2)=10^5$ $cm^{-3}$. In addition, the variation 
of the power-law index along the jet suggests that the portion of hot gas 
decreases with distance from the jet engine.  The 
molecular outflow in this region has been mapped for the first time using 
CO data.  
From the calculated outflow momentum flux, a very 
strong lower limit to the average accretion luminosity is 
$3.6 \, \frac{\mathrm{sin} \, i}{\mathrm{cos}^3 \, i}$ \lsun, 
indicative of a decrease in the accretion rate with time.    

\end{abstract}


\section{INTRODUCTION}

Stars form out of the collapse of dense molecular gas cores.  However, this 
process is complex, and the details are far from understood.
Young stellar objects (YSOs) and their immediate
surroundings are expected to evolve significantly throughout star formation.
For example, YSOs will evolve in luminosity and temperature, as
their masses grow and they evolve toward the main sequence 
(e.g., Myers et al. 1998).  In addition, the
associated gas will evolve dynamically.  As collapse proceeds gas flows from
cores through disks and onto protostars, driving jets and outflows in the 
process.  These outflows are a
crucial mechanism by which angular momentum is removed from a star forming 
system where accretion occurs through a rotating disk.  Therefore, an 
outflow is a good tracer
of star formation, especially when the central protostar is deeply embedded.

Large molecular outflows have been mapped in (sub)millimeter lines of CO 
(Fukui 1989, Morgan et al. 1991), which trace the matter cooled down after 
being shocked.  However, more recently shocked material such as warm 
molecular hydrogen produces ro-vibrational or pure rotational transitions 
at infrared wavelengths. The Infrared Spectrograph 
(IRS) aboard the $Spitzer$ Space Telescope (SST) has been used to map 
the shocked H$_2$ line emission (Neufeld et al. 2006, Maret et al. 2009). 
Lately, Neufeld \& Yuan (2008) 
showed that the {\it Spitzer} Infrared Array Camera (IRAC) 
intensities were dominated by warm molecular hydrogen emission in 
the supernova remnant IC 443.  
In addition, Ybarra \& Lada (2009) also showed that IRAC 
colors could be used to study properties of the hot, shocked gas components 
in HH54. 
 
The $Spitzer$ Legacy Program ``From Molecular Cores to Planet 
Forming Disks" (c2d; Evans et al.\ 2003) discovered a new jet feature
on the western edge of L1251 (core A in Sato et al. 1994), 
whose distance is $300(\pm 50)$ pc (Kun \& Prusti 1993).  
No outflows were previously detected in this region although 
there are three IRAS sources (Kun \& Prusti 1993; Sato et al. 1994).  
No optical jets or Herbig Haro objects have been reported in this 
region, probably due to the high optical depth. 
The discovered infrared jet is the longest ($\sim 1$ pc) except HH 111, which 
is onesided and $\sim 8$ pc (Reipurth 1989, Reipurth et al. 1997).  
In this paper, we examine the properties of warm molecular gas along the jet 
feature with the IRAC images.  We also study the molecular outflow and the 
embedded YSO associated with the jet using, respectively, CO data and a set of 
infrared through millimeter photometric data.

\section{OBSERVATIONS}

The $Spitzer$ c2d Legacy program observed core A of L1251 (L1251-A) 
at 3.6, 4.5, 5.8, and 8.0 \um\ with the Infrared Array Camera 
(IRAC; Fazio et al. 2004) on 4 September 2004 and at 24 and 70 \um\ with 
the Multiband Imaging Photometer for $Spitzer$ (MIPS; Rieke et al. 2004) 
on 30 November 2004 (PID:139;  AOR keys: IRAC 5165824 and MIPS 9425664).
The detailed description of the data processing can be found in the c2d final
delivery documentation (Evans et al. 2007).

To cover the whole region of the infrared jet, the CO 2$-$1 transition 
at 230.537970 GHz has been mapped with the 6 m telescope 
at the Seoul National Radio Astronomical Observatory in March and April 2009.
The beam FWHM is 48\as\ at 230 GHz.  The main beam efficiency and pointing 
accuracy are 0.57 and $\sim 3\as$, respectively.  The velocity resolution 
is 0.127 km s$^{-1}$ after binning by two channels.  

For the analysis of the Spectral Energy Distributions (SEDs), we also use 350 \um\ and 1.2 mm continuum emission maps observed with  the
Submillimeter High Angular Resolution Camera II (SHARC-II) mounted on the
Caltech Submillimeter Observatory, and the Max-Planck Millimeter Bolometer (MAMBO) at the IRAM 30m telescope, respectively.
Details of these continuum data can be found in Wu et al. (2007) and Kauffmann et al. (2008).

\section{RESULTS}

Figure 1a shows a three-color image comprised of IRAC 3.6, 4.5, and 8.0 \um\ 
images. The dominant feature in the image is a jet extending about 10\am.  
The extended blue color emission perpendicular to the jet is consistent with 
the distribution of the millimeter dust continuum emission (see upper panel of 
Figure 1b), so it is likely dominated by scattered light within a dense 
cloud with some contribution from shocked gas associated with jets and 
outflows.  There is also a small bipolar nebula structure, similar to 
that seen in core E of L1251 (L1251B, Lee et al. 2006), to the east of the long jet.   

In Figure 1a, stellar objects (IRS3 and IRS4) associated with the jet and 
bipolar nebula are not clearly seen.  However, these central engines of 
the jet and bipolar structure are well detected at 24 and 70 \um\ (see Table 1 
for fluxes).  At 70 \um, the two infrared sources become brighter (see 
upper panel of Figure 1b).  
In addition, as seen in Figure 1b, these sources are each located at both 
350 \um\ and 1.2 mm dust continuum peaks, indicating they are likely very 
embedded objects.  
In L1251-A the c2d team identified 4 YSOs, marked as IRS1, IRS2, IRS3, and IRS4 
in Figure 1b.  The infrared source between IRS3 and IRS4 is classified as a 
galaxy based on the c2d criteria (Evans et al. 2007).  Table 1 lists fluxes of 
the four YSOs at various wavelengths, and the SED of the YSO associated with 
the jet (IRS3) is shown in Figure 2 (as red diamonds).  The SED clearly 
indicates that IRS3 has a very thick envelope with strong (sub)millimeter 
emission.  IRS4 is associated with the bipolar nebula and shows a similar 
SED to that of IRS3, although it is not presented in this paper.  

Figure 1b (lower panel) shows the CO 2$-$1 outflow map on top of the IRAC 
4.5 \um\ image.  The outflow map covers the whole jet feature as well as 
the bipolar nebula.  Along the jet, it shows two distinct lobes with little 
overlap, suggesting a large inclination of rotational axis with respective to
the line of sight.  
The outflow lobes are well 
correlated spatially with the infrared jet.  No clear outflow structure is 
seen towards IRS4 and its associated bipolar nebula, although some weak 
red-shifted emission is present.

\section{ANALYSIS}

The jet feature associated with IRS3 appears consistently in all IRAC bands, 
indicating that the jet emission is probably produced by a common gas 
component such as molecular hydrogen, which has its line transitions 
distributed over all IRAC bands.  Therefore, we calculate the IRAC colors of 
[5.6 \um]/[8.0 \um] and [4.5 \um]/[8.0 \um] to compare with those of thermal 
gas modeled to explain the molecular hydrogen emission from the jet 
(Figure 3).  The colors were calculated in 4 pixel ($\sim 5\as$) radius 
apertures at the positions marked with $black$ circles in Figure 2.  
In the model calculation of the colors (Figure 3), the molecular gas was 
assumed to be composed of H$_{2}$ and He, with $ n(He)=0.2~n(H_{2})$.  We 
calculated the level population of H$_2$ gas in statistical equilibrium. The 
collisional coefficients were obtained from Le Bourlot et al. (1999), and an 
ortho-to-para ratio of 3.0 is assumed.  The temperature of the shocked H$_{2}$ 
gas was modeled in two ways: isothermal and power-law admixed. In the latter, 
the infinitesimal H$_{2}$ column density has a power-law relation with the 
temperature $T$, d$N\sim T^{-b}dT$, in the range of $300 \sim 4000$ K.  
More detailed descriptions can be found in Neufeld \& Yuan (2008) and Shinn et 
al. (2009a, b).  

We know that this jet is located inside a molecular cloud (Sato et al. 1994), 
thus, 
the extinction close to the jet engine is possibly much higher than 0.5 mag, 
which is the interstellar extinction toward L1251-A (Kun \& Prusti 1993).
Therefore, the expected colors are extinction corrected with A$_V\sim 5$ mag, 
which is obtained from the 1.2 mm dust continuum emission
at 0.1 pc from IRS3, following the ``Milky Way, $R_V=3.1$" curve 
(Weingartner \& Draine 2001, Draine 2003). 
Extinction corrected colors with A$_V\sim 0.5$ and 5 mag are not different.
However, close to YSOs (C and E in Figure 1a), A$_V\sim 50$ mag. 
We tested this high extinction only for the colors of C and E, which
move diagonally up and right in the color-color diagram.

According to the comparison between the observed and modeled colors along the 
jet, the IRAC colors can be explained with the power-law admixture model, as 
Neufeld \& Yuan (2008) and Shinn et al. (2009a, b) found.  All colors are 
located between the model of $n(H_2)=10^6$ cm$^{-3}$ and the LTE line, where 
the pre-shock density must be higher than $10^5$ cm$^{-3}$ (Wilgenbus et al. 
2000).  There might be some contribution from other emission components not 
considered in our model, such as the CO $v=1-0$ transitions in IRAC 2.  
However, such contamination is not likely according to Neufeld \& Yuan (2008).  
The power-law index, 
$b$ is the smallest ($\sim 2$) at the jet feature close to the central 
embedded source (C in Figure 1a) and the bipolar nebula feature (E in Figure 
1a) in the east of the long jet, indicating that there are more hot components 
close to the embedded YSOs.  Alternatively, the scattered light in the outflow 
cavity may enhance the colors (more scattered light at shorter wavelengths).  
If we apply A$_V=50$ mag, the colors at C and E become similar to those of
the model with b$\sim 1$.
Except for these two points, $b$ is between $3.3-5.2$, consistent with results 
of Neufeld \& Yuan (2008), who showed that this range of $b$ values can be 
explained by the paraboloidal bow shock model.  Bow shocks can be considered 
as a geometrical summation of planar C-shocks, whose postshock temperature is 
nearly isothermal (Neufeld et al. 2006). In this sense, Neufeld \& Yuan (2008) 
showed that a single bow shock generates a shocked H$_2$ gas with b$\sim 3.6$, 
and $b$ can be higher than 3.8 when slow bow shocks, where H$_2$ is not 
dissociated even at the apex, are mixed with fast bow shocks.

Following Young et al. (2004), Bourke et al. (2006), and Dunham et al. (2006), 
we modeled the SED of the YSO associated with the jet (IRS3) using the 1-D 
dust continuum radiative transfer code DUSTY (Ivezic et al. 1999).  In 
addition to a central stellar object, a 1-D disk model (Butner et al. 1994) 
was adopted with a surface density profile of $\Sigma (r) \propto r^{-1.5}$ 
and temperature profile of $T(r)\propto r^{-0.35}$ to calculate the internal 
input SED for the best-fit model.  The envelope is assumed to have a power-law 
density profile and is heated by the interstellar radiation field attenuated 
by A$_V$=3 mag.  For the dust opacity, we adopt the same dust model used by 
Crapsi et al. (2008): a mixture of carbon (29\% of the total dust mass) and 
silicate grains covered with various molecular ices.  For a detailed 
explanation of the modeling procedure see Young \& Evans (2005) and Dunham et 
al. (2006).  The best-fit SED to the observed fluxes is presented in Figure 2. 
The best-fit model has an internal luminosity of $\sim 0.9$ \lsun\ with 
$T_{star}=2500$ K.  
This internal luminosity derived by the model, which is considered to be the 
true bolometric luminosity of the source, is greater than \lbol\ calculated 
from observed fluxes because some of these long-wavelength fluxes were measured 
with apertures smaller than the total extent of the source.  The envelope of 
the model has a power-law density profile index of $p=1.55$, a mass of $\sim 
12$ \msun, and 
inner and outer radii of 250 and 18,000 AU, respectively.
The internal luminosity derived by the model can be considered the current 
accretion luminosity. 

The average accretion luminosity over the YSO lifetime can be derived from the 
momentum flux in the outflow.  We use the CO J$=2-1$ map to calculate the 
outflow momentum flux, $F_{\rm CO}$, following the method presented by 
Hatchell et al. (2007).  This quantity is the average rate at which momentum 
is injected into the outflow, and we calculate $F_{\rm CO} \geq 5.9 \times 
10^{-6}$ $\frac{\mathrm{sin} \, i}{\mathrm{cos}^2 \, i}$ \msun\ \kms\ 
yr$^{-1}$ and list this result in Table 2.  There is an inclination dependence 
that we leave in the result since the source inclination is not known.  Two 
additional unknowns in the above calculation are the optical depth and the 
temperature of the outflowing gas (needed to convert line intensity to column 
density assuming LTE).  Following M. Dunham et al. (2009, in prep), we assume 
the gas is optically thin and has a temperature of $17.6$ K, which minimizes 
the line intensity to column density conversion.  Optically thick gas and any 
other temperature in the range of $10-100$ K will increase $F_{\rm CO}$, thus 
the above result is a very strong lower limit to the true value.

Since molecular outflows are driven by the transfer of momentum from a 
jet/wind ejected by the protostellar system to the ambient medium and the 
jet/wind ejection process is closely tied to accretion onto the protostar, the 
calculated $F_{\rm CO}$, which represents the average rate at which momentum 
is injected into the outflow, is related to $\langle \dot{M}_{\rm acc}\rangle$, 
the time-averaged protostellar mass accretion rate (see Bontemps et al. 1996):

\begin{equation}
\langle\dot{M}_{\rm acc}\rangle = \frac{1}{f_{\rm ent}} \, 
\frac{\dot{M}_{\rm acc}}{\dot{M}_w} \frac {1}{V_w} \, F_{\rm CO} 
\quad .
\end{equation}
In the above equation,  $\dot{M}_w$ is the mass-loss rate in the jet/wind, 
$V_w$ is the velocity of the jet/wind, and $f_{\rm ent}$ is the entrainment 
efficiency between the jet/wind and the ambient gas.  With typical values for 
the above parameters of $ \frac{\dot{M}_{w}}{\dot{M}_{\rm acc}} =0.1$, 
$V_w\sim 150$ \kms, and $f_{\rm ent}=0.25$ (Bontemps et al. 1996 and 
references therein), we calculate $\langle \dot{M}_{\rm acc} \rangle \geq 1.6 
\times 10^{-6} \frac{\mathrm{sin} \, i}{\mathrm{cos}^2 \, i}$ \msun\ 
yr$^{-1}$.  An outflow dynamical time ($\tau_d \geq 14.2 \times 10^4 \, 
\frac{\mathrm{cos} \, i}{\mathrm{sin} \, i}$ yr) is calculated by dividing the 
extent of the outflow ($400\as/\mathrm{sin} \, i$) by the outflow velocity 
($\sim 4/\mathrm{cos} \, i$ \kms).  Assuming accretion at the above rate for 
this length of time gives an accreted protostellar mass of $M_{\rm acc} \geq 
0.22 \, \frac{1}{\mathrm{cos} \, i}$ \msun.  Finally, accretion at the above 
rate onto a protostar with this mass and $R=3$ \rsun\ gives a spherical 
accretion luminosity of $L_{acc} = \frac{GM\langle\dot{M}_{acc}\rangle}{R} 
\geq 3.6 \, \frac{\mathrm{sin} \, i}{\mathrm{cos}^3 \, i}$ \lsun.

\section{DISCUSSION}

We calculated the bolometric luminosity and temperature of each YSO 
identified in L1251-A, following the method used in Dunham et al. (2008) 
(see Table 1).  The apparent drivers (IRS3 and IRS4) of the infrared jet and
bipolar nebula have \lbol\ $\le 1$ $L_\odot$, similar to IRS2 in L1251B 
(L1251B-IRS2), another source associated 
with infrared bipolar structure (Lee et al. 2006).  Although IRS3 and 
IRS4 are $\sim 10$ times fainter than L1251B-IRS2 at 24 \um, they are very 
bright at wavelengths longer than 70 \um.  Their bolometric temperatures 
are $\sim$20 K, much lower than that of L1251B-IRS2 (140 K).  
Finally, the ratio of $L_{smm}/L_{bol}$ for IRS3 and IRS4 ($\sim$0.13 
and 0.17, respectively, where $L_{smm}$ is the luminosity at $\lambda > 350$ 
\um) is much greater than that of L1251B-IRS2 ($\sim$0.03), indicative of
Class 0 sources with very thick envelopes.

The dominant feature in L1251-A is the long, well-collimated jet associated 
with IRS3.  This type of jet structure can be produced in the IRAC bands by 
ro-vibrational or pure rotational line emission of H$_2$ in a pulsating jet 
with little precession (model A1 in Smith \& Rosen 2005).  According to Smith 
\& Rosen (2005), this jet model (A1) transfers the bulk kinetic energy to 
large distances but excites H$_2$ with the lowest efficiency.  
However, the jet shows undulating structure toward both ends. Smith \& Rosen
(2005) showed this kind of helical stream in a jet with a slow uniform-speed 
precession. Therefore, the jet in L1251-A might be explained with a small 
precession angle (but greater than that of A1) and a long pulsating period 
(much longer than 60 years). 
In fact, the jet associated with the northern, red component of the CO outflow 
features two distinct emission bulks, indicative of episodic ejection.  
The time interval between episodes of ejection is $1300/{\mathrm{sin} \, i}$ 
years if we assume a jet speed of 150 \kms. 
 
The infrared jet and the CO outflow are well correlated in shape and length
as seen in Figure 1b. From the CO $2-1$ map, we calculated 
a time-averaged mass accretion rate and the accretion luminosity expected 
from accretion at this rate (Table 2).  
In spite of uncertainties in $f_{\rm ent}$ and 
${\dot{M}_{w}}/{\dot{M}_{\rm acc}}$, the calculated time-averaged accretion 
rate (and resulting accretion luminosity) are strong lower limits, as explained 
in \S 4.
In addition, a large inclination (there is little 
overlap between red and blue components of the CO outflow and no
significant intensity difference between two infrared jet lobes) of 
70\degree\ will give  $L_{acc}\sim 83$ \lsun, which is much greater 
than $L_{acc}$ derived from dust modeling (0.9 \lsun).
Since the luminosity from dust modeling depends on the \emph{current} 
mass accretion rate, this rate must be lower now than the time-averaged value 
over the outflow lifetime.

The comparison between the observed IRAC colors and the ones from the model of 
a power-law admixture of gas (see Figure 3) indicates 
the amount of hot components varies 
(in general, more hot gas closer to the jet engine).  If we use the same 
colors ([3.6]-[4.5] vs. [4.5]-[5.8]) as those presented by Ybarra \& Lada 
(2009), who modeled only the hot (2000$-$4000 K) gas component, C and E 
($b\sim 2$, see Figure 3) are located at $n(\rm H) > 10^5$ cm$^{-3}$, 
while N2, N3, and S ($b\sim 3.3-4.5$) lie close to the model line of $n(\rm H) 
= 10^5$ cm$^{-3}$.  Therefore the density inferred by 
their model is not very different from that inferred from our model, 
although, unlike in our model, Ybarra \& Lada (2009) included collisions 
between H$_2$ and H.  However, N1 is placed at $n(\rm H) \sim 2\times 10^3$ 
cm$^{-3}$ in their model, consistent with our model result since the model of 
Ybarra \& Lada (2009) considered only the hot gas component, and in our model, 
$b\sim 5.2$ for N1 suggests a relatively small amount of hot gas at that
position.    

\acknowledgments
We acknowledge the support by the Korea Science and Engineering Foundation 
(KOSEF) through the Astrophysical Research Canter for the Structure and 
Evolution of the Cosmos (ARCSEC).
This work was also supported in part by a grant (R01-2007-000-20336-0) 
from the Basic Research Program of the KOSEF.
Support for this work, part of the Spitzer Legacy Science
Program, was provided by NASA through contract 1224608 
issued by the Jet Propulsion Laboratory, California Institute of
Technology, under NASA contract 1407.


\clearpage

\begin{deluxetable}{ccrrrrrrrrrcccc}
\tabletypesize{\scriptsize}
\rotate
\tablecolumns{14}
\tablewidth{0pc}
\tablecaption{Properties of YSO candidates in L1251-A}
\tablehead{
\colhead{} & \colhead{} & \colhead{} & \colhead{} & \multicolumn{7}{c}{FLUXES (mJy)} \\
\cline{4-11}\\ 
\multicolumn{1}{c}{YSOs} & \colhead{R.A.} & \colhead{DEC.} &
\colhead{3.6 \um} & \colhead{4.5 \um} & \colhead{5.8 \um} & \colhead{8.0 \um} &\colhead{24 \um} 
& \colhead{70 \um} & \colhead{350 \um\tablenotemark{a}} & \colhead{1.2 mm\tablenotemark{b}} 
& \colhead{$L_{bol}$ (\lsun)} & \colhead{$T_{bol}$ (K)} & \colhead{\fsmm}}
\startdata
IRS1\tablenotemark{c} & 22 29 33.4 & 75 13 15.9 & 5.62 & 10.50 &
 15.60 &      18.20 &      55.8 &      150& - & - & 0.07 & 290 & -\\
IRS2\tablenotemark{d} & 22 29 59.5 & 75 14 03.2 & 15.80 & 21.80 &      
 25.20 &      27.30 &   272 &      848& - & 713& 0.4 & 230 & -\\
IRS3\tablenotemark{e} &  22 30 31.9 & 75 14 08.8 & 0.13 & 0.42 &     
 0.32 &     0.16 &      4.97  &      1400& 9400 & 1020& 0.8 & 24 & 0.12\\
IRS4\tablenotemark{f} & 22 31 05.6 & 75 13 37.1 & 0.43 & 1.07 &     
0.75 &     0.37 &      1.90 &      688& 7500 & 925 & 0.6 & 21 & 0.13\\
\enddata
\tablenotetext{a}{(SHARCII) aperture size = 40 \as}
\tablenotetext{b}{(MAMBO) aperture size = 80 \as}
\tablenotetext{c}{SSTc2d J222933.4+751316}
\tablenotetext{d}{SSTc2d J222959.5+751403 (IRAS 22290+7458)}
\tablenotetext{e}{SSTc2d J223031.9+751409}
\tablenotetext{f}{SSTc2d J223105.6+751337}
\end{deluxetable}

\begin{deluxetable}{lll}
\tabletypesize{\scriptsize}
\tablewidth{0pt}
\tablecaption{\label{tab_outflow}Outflow and Mass Accretion Properties}  
\tablehead{\colhead{Quantity} & \colhead{Units} & \colhead{Value}}
\startdata
Outflow Momentum Flux ($F_{\mathrm CO}$) & $10^{-6}$ \msun\ \kms\ yr$^{-1}$ & $\geq$ 5.9 $\frac{\mathrm{sin} \, i}{\mathrm{cos}^2 \, i}$\tablenotemark{a} \\
Average Protostellar Mass Accretion Rate ($\langle \dot{M}_{\mathrm acc} \rangle$) & $10^{-6}$ \msun\ yr$^{-1}$ & $\geq$ 1.6 $\frac{\mathrm{sin} \, i}{\mathrm{cos}^2 \, i}$ \\
Dynamical Time ($\tau_{\mathrm d}$) & 10$^4$ yr & $\geq$ 14.2 $\frac{\mathrm{cos} \, i}{\mathrm{sin} \, i}$ \\
Protostellar Mass Accreted ($M_{\rm acc}$) & \msun & $\geq 0.22 \, \frac{1}{\mathrm{cos} \, i}$ \\
Accretion Luminosity $\,$ (\lacc) & \lsun & $\geq 3.6 \frac{\mathrm{sin} \, i}{\mathrm{cos}^3 \, i}$  \\
\enddata
\tablenotetext{a}{$i$ is the inclination of the rotational axis with respect to the line of sight.}
\end{deluxetable}

\clearpage

\begin{figure}
\figurenum{1}
\epsscale{1.1}
\hskip -2cm
\plotone{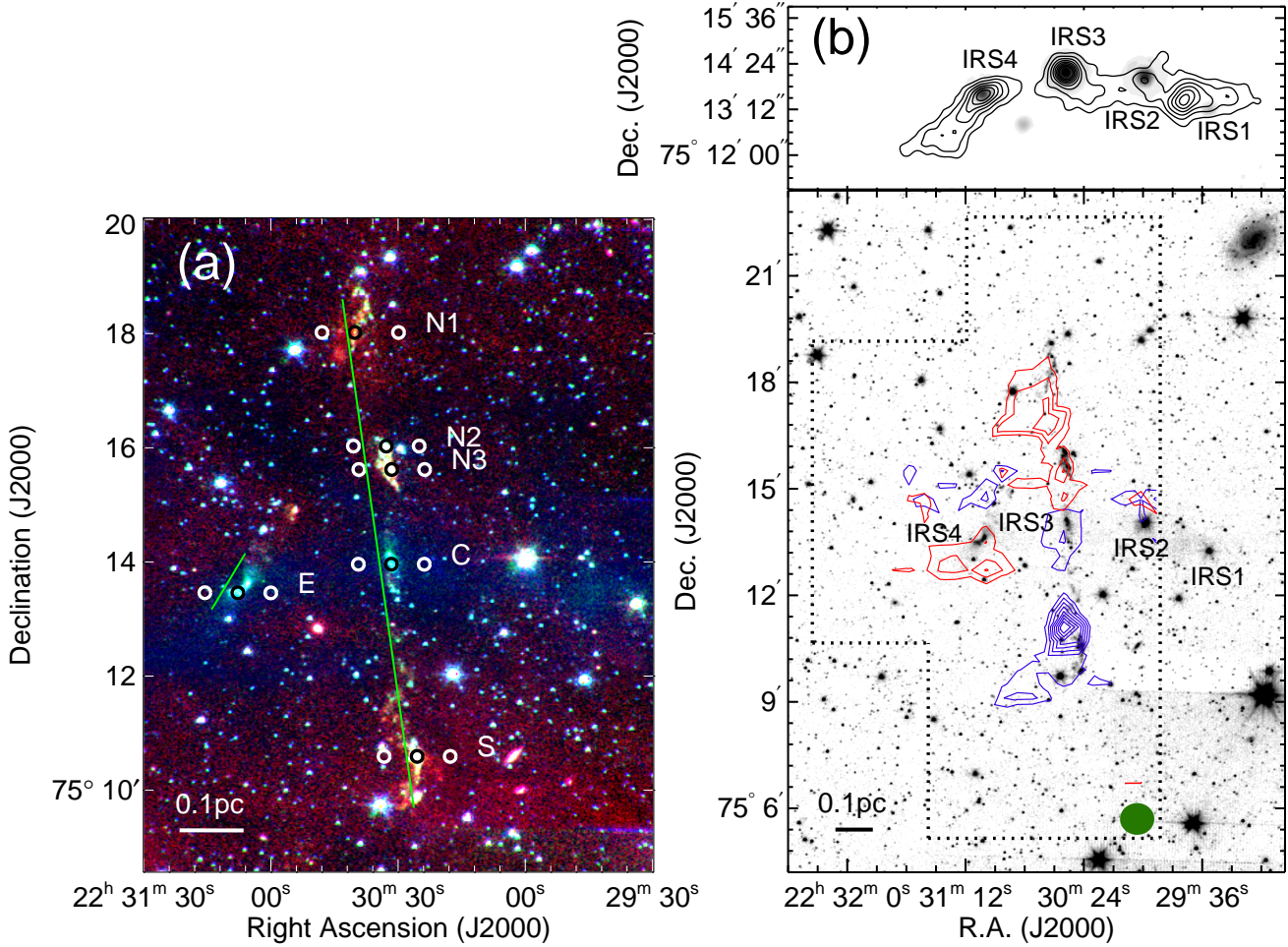}
\caption{{\bf (a)} Three-color composite $Spitzer$ image of L1251-A. 
The IRAC 3.6, 4.5, and 8.0 \um\ data are
presented, respectively, as blue, green, and red. Green lines are placed along 
the jet and bipolar nebula features. $Black$ circles show the regions where 
the colors plotted in Figure 3 were derived, and $white$ circles to the east 
and west of each $black$ circle indicate where the background was measured. 
{\bf (b)} {\bf\it Upper Panel:} 
1.2 mm MAMBO intensity map (contours) on top of MIPS 70 \um\ image.  
IRS3 and IRS4 are located at the peaks of MAMBO emission. 
Another emission peak exists between IRS1 and IRS2, which might be a 
prestellar core. 
{\bf\it Lower Panel:} CO molecular outflow map (contours) on top of IRAC 4.5 
\um\ image. 
The dotted line encloses the total area mapped, and the green filled circle
at the right bottom denotes the beam at 230 GHz.
The blue contours are integrated from -13.0 to -7.0 \kms, while the red 
contours are 
integrated from -1.0 to 5.0 \kms. 
The contours start at 1.2  and 1.6 K \kms\ for blue and red components, 
respectively, and increase by 0.4 K \kms. 
The lobes of the main outflow are well correlated with 
the infrared jet along the NS direction. 
}
\end{figure}

\clearpage

\begin{figure}
\figurenum{2}
\epsscale{1.0}
\plotone{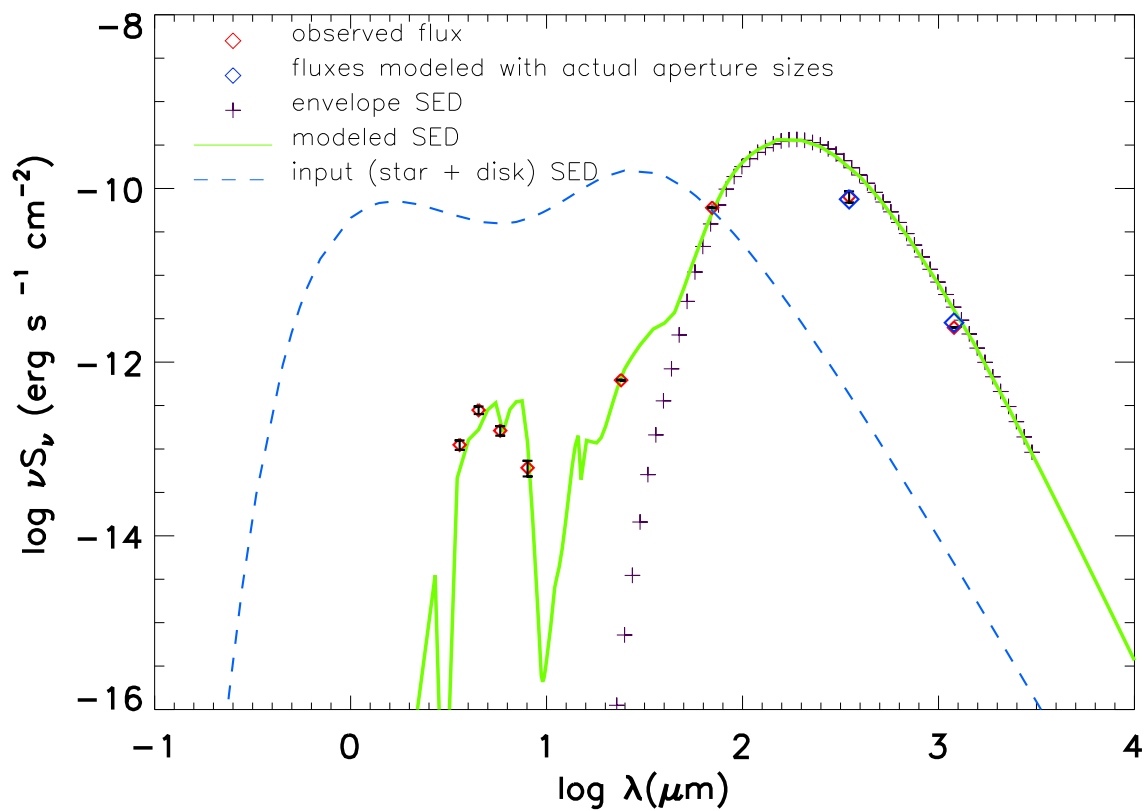}
\caption{
The best-fit model SED (green solid line) to the observed fluxes 
(red diamonds) of IRS3, with the internal input SED shown (blue dashed line). 
Blue diamonds represent model fluxes in the apertures used for photometry.  
Crosses indicates the SED emerging only from the envelope. This SED modeling 
gives an internal luminosity for IRS3, produced mostly by accretion, 
of $\sim 0.9$ \lsun.
}
\end{figure}

\begin{figure}
\figurenum{3}
\includegraphics[scale=0.8,angle=90]{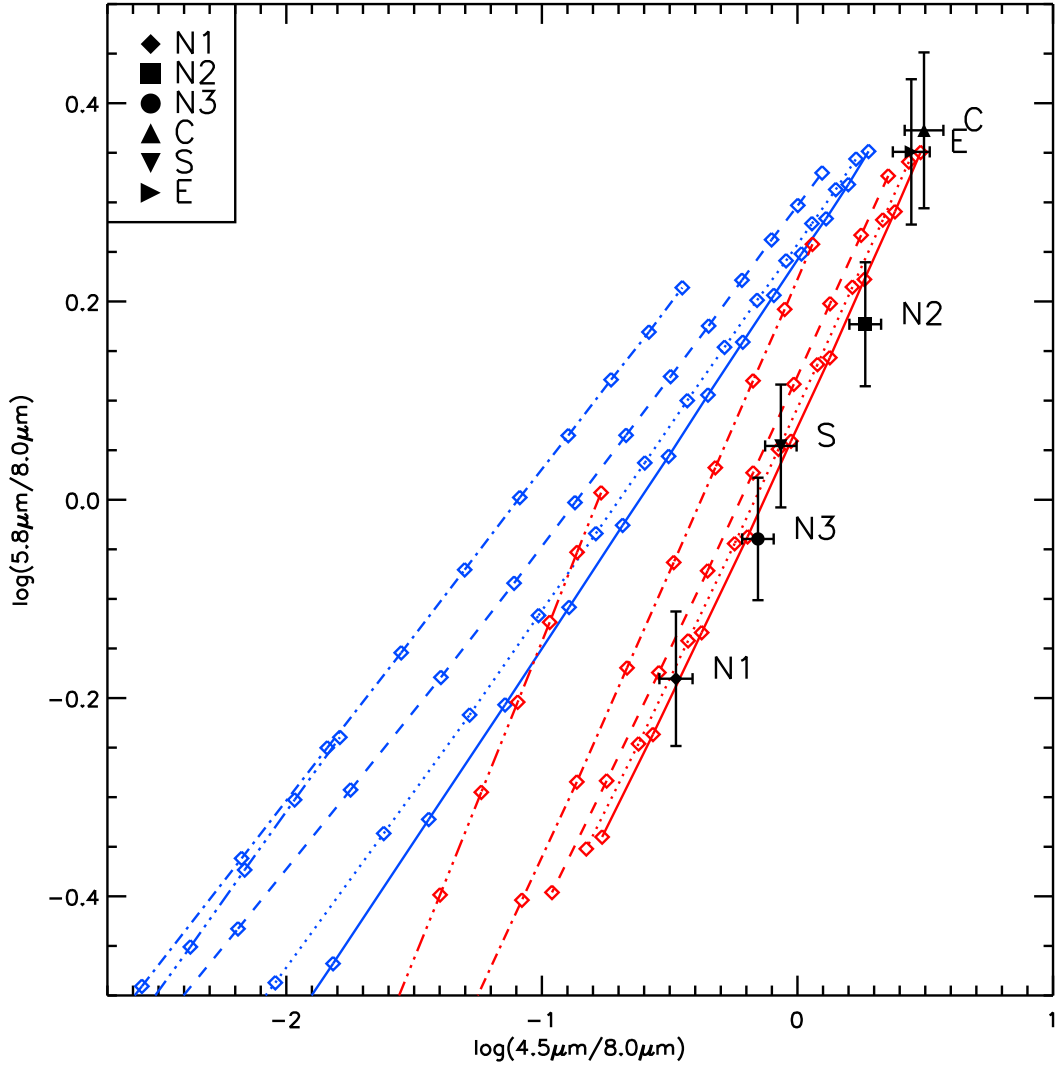}
\caption{Two-color diagram of [4.5 \um]/[8.0 \um] and [5.6 \um]/[8.0 \um] 
expected from pure H$_2$ emission with an assumption of $A_v=5$ mag. 
{\it Blue lines}: single-component models for n(H$_2$)$=10^4$ 
(dot-dot-dot-dashed), 
$10^5$ (dot-dashed), $10^6$ (dashed), $10^7$ cm$^{-3}$(dotted), and LTE 
(solid). Open squares represent the results of a range of temperatures up to 
2000 K ({\it top right of each curve}) in steps of 100 K. {\it Red lines}: 
models with a power-law temperature distribution, with $dN=aT^{-b}dT$ and $b$ 
in the range of 2 ({\it top right of each curve}) to 6. The line type for each 
model is the same as 
in the single-component models.  Each symbol represent the color measured at 
each position marked in Figure 1a. The error bars include the flux measurement 
error and the flux calibration error ($\sim 10\%$).
}
\end{figure}

\end{document}